\documentclass[epj,referee]{svjour}
\usepackage{graphics}
\begin{document}

\title{A new view of the electronic structure of
the spin-Peierls compound $\alpha'$-NaV$_2$O$_5$}

\author{P. Horsch and F. Mack \inst{ }}
\institute{
Max-Planck-Institut f\"{u}r Festk\"{o}rperforschung,
Heisenbergstr.~1, D-70569 Stuttgart (Germany) }

\date{January 28, 1998}

\abstract{
The present understanding of the electronic and magnetic properties of
$\alpha'$-NaV$_2$O$_5$ is based on the hypothesis of strong charge
disproportionation into V$^{4+}$ and V$^{5+}$, which is assumed to lead
to a spin-1/2 Heisenberg chain system. 
A recent structure analysis shows, however, that the V-ions are in a mixed
valence state and indistiguishable. We propose an explanation for
the insulating state, which is not based on charge modulation, and
show that strong correlations together with 
the Heitler-London character of the relevant intermediate
states naturally lead to antiferromagnetic Heisenberg chains. 
The interchain coupling is weak and frustrated, and its effect on the
uniform susceptibility is small.
\newline
{\it Dedicated to J. Zittartz on the occasion of his 60th birthday}
\PACS{{75.50.Ee}{ }\and
      {75.30.Et}{ }\and
      {75.40.Cx}{ }\and
      {75.40.Mg}{ } }  
}

\maketitle

One-dimensional spin-1/2 Heisenberg antiferromagnets are expected to
undergo a structural phase transition into a dimerized phase
at low temperature 
accompanied by the opening of a spin gap\cite{Pytte74}.
This spin-Peierls transition was first observed in organic systems
\cite{Bray75}, but has found considerable experimental attention
after its recent discovery in CuGeO$_3$ ($T_{sp}=14 K$)\cite{Hase93}. 
The $\alpha'$-phase of NaV$_2$O$_5$ appears to be
the second inorganic compound where a similar transition was
observed with an even higher transition temperature $T_{sp}=34 K$
\cite{Isobe96}.
The size of the spin gap was determined by neutron scattering
\cite{Fujii97}, Raman\cite{Weiden97} and other experiments.  
The transition into the low-temperature dimerized structure was 
confirmed by X-ray 
scattering\cite{Fujii97}, NMR\cite{Ohama97},
thermal-expansion\cite{Koeppen98}
yet the detailed deformation pattern is still unknown.

Based on an early structure determination for  $\alpha'$-NaV$_2$O$_5$
\cite{Carpy75},
the current picture for the origin of the one-dimensional magnetic
properties rests on the assumption of charge discommensuration into
V$^{4+}$ and  V$^{5+}$ chains\cite{Isobe96,Fujii97,Ohama97,Mila96,Augier97}. 
In Fig.1 the V1 and V2 chains in
b-direction would correspond to these different charged vanadium
chains. In this picture the V$^{4+}$ chain would correspond to a
spin-1/2 Heisenberg chain.
A problem with this picture is, however, the missing physical argument
for such a strong charge modulation. 

In fact a recent new structural analysis of $\alpha'$-NaV$_2$O$_5$
by von Schnering {\it et al}\cite{Schnering97}
clearly shows that all vanadiums are equivalent and in a mixed valent state. 
Furthermore the charge-ordered state at room temperature was found to
be in conflict with Raman scattering experiments\cite{Fischer98}.
Since there is one d-electron per two V ions an explanation for the
insulating properties must be found which does not rest on charge
discommensuration.  

The aim of this work is to show how in such an intermediate valence
situation correlation effects can lead to a quasi-1D Heisenberg
antiferromagnet. An important structural information\cite{Schnering97},
which is crucial
for our analysis, are the V-O distances $d$ in the a-b plane (Fig.1):
$d$(V1-O3)=$d$(V2-O3)=1.825$\AA$, $d$(V1-O2$\parallel$b)=1.916$\AA$
and $d$(V1-O2$\parallel$a)=1.986$\AA$. This implies that the V1-O3-V2
bond is much shorter than the V1-O2-V1 bond in b-direction. 
This additional bonding effect can be attributed to the
$d_{xy}$-electrons. From Fig. 1 we see that the $d_{xy}$-orbitals have
a direct overlap in one direction, e.g. V2-V1$'$, which is however quite
small $t_{xy}\cong 0.3$ eV\cite{Grin97}. 
Yet a $d_{xy}$-electron can also hop
between V1 and V2 via a double exchange like process. 
In such a process first an
electron hops from the occupied O3 $p_y$-orbital to V2 with an
excitation energy $\Delta\epsilon_y =\epsilon(d_{xy})-\epsilon(p_y)$,
and in a second step the V1-electron annihilates the oxygen hole. 
The matrixelement for this process is $t_a\cong t^2_{pd}/\Delta\epsilon_y$.
The matrixelement $t_{pd}$ depends on the vanadium-oxygen distance $d$
and can be estimated with the help of Harrison's\cite{Harrison80}
solid state table as $t_{pd}=\eta_{pd\pi}\hbar^2r_d^{3/2}/md^{7/2}$, where
$\eta_{pd\pi}=1.36$, $\hbar^2/m=7.62$ eV$\AA^2$, and $r_d=0.98\AA$ for vanadium.
This gives $t_{pd}\cong 1.2$ eV.
\begin{figure}
\resizebox{0.40\textwidth}{!}{%
\includegraphics{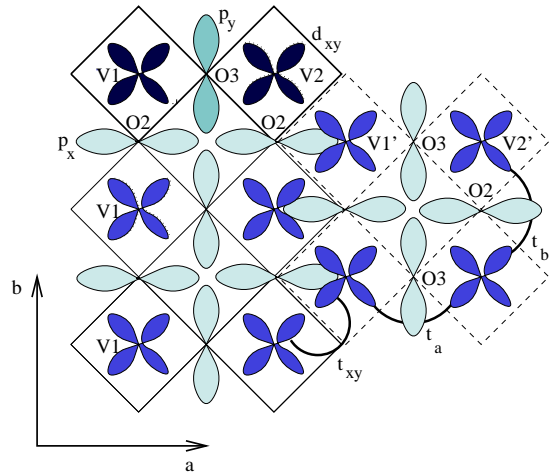}
}
\vspace{10mm}
\noindent
\caption{Orbital structure of $\alpha'$-NaV$_2$O$_5$ in the a-b plane.
  The lowest d-orbitals with $d_{xy}$-symmetry are occupied by one
  electron per two V-atoms. The O p-orbitals at the corners of square
  pyramids (where only the basal plane is indicated) are occupied, with
  energies ranging from -3 eV down to -7 eV. 
  Solid (dashed) squares indicate the downward (upward) orientation
  of the pyramids (for structural details see Refs.[12] and [4]). 
  The $d_{xy}$ orbitals
  have a direct overlap $t_{xy}$ along the V2-V1$'$-V2-V1$'$ zig-zag
  chain. The largest hopping matrix element $t_a$ is, however, via
  double exchange interaction involving an O3 $p_y$ orbital. The
  resulting bonding is manifested in the structure by the shortest
  planar V-O distance along the V1-O3-V2 bond.
\label{fig1}}
\end{figure}
The $p$-$d$ excitation energies are
determined from a LMTO-bandstructure calculation\cite{Grin97}, which yields
$\Delta\epsilon_y\cong 4$ eV and $\Delta\epsilon_x\cong 6.5$ eV,
respectively. Hence $t_a\cong 0.35$ eV, while the corresponding
matrixelement $t_b\cong 0.15$ eV is considerably smaller because of the
larger oxygen-vanadium distance and  the larger $\Delta\epsilon_x$.
We note that the larger value for $t_a$ is consistent with the
stronger bonding of the  V1-O3-V2 bond.

The Hamiltonian for the $d_{xy}$-electrons may be written in terms  
of creation and density operators $d^{\dagger}_{{\bf i}\alpha\sigma}$ and
$n_{{\bf i}\alpha\sigma}=d^{\dagger}_{{\bf i}\alpha\sigma} d_{{\bf
    i}\alpha\sigma}$, respectively, in the form 
\begin{eqnarray}
H_d =&-&\sum_{{\bf i}\sigma}t_a (d^{\dagger}_{{\bf i}1\sigma} d_{{\bf
    i}2\sigma} + H.c.)
    + U_d\sum_{{\bf i}\alpha}
    n_{{\bf i}\alpha\uparrow} n_{{\bf i}\alpha\downarrow}   \nonumber\\
   &-&\sum_{\langle {\bf i j} \rangle \alpha\beta\sigma} t_{\bf ij} 
   (d^{\dagger}_{{\bf i}\alpha\sigma} d_{{\bf j}\beta\sigma} + H.c.),
\label{eq:H}
\end{eqnarray}
where we have introduced a cell structure. Here a cell contains two
vanadium atoms V1 and V2, i.e. $\alpha=1$ and $2$, and is labeled by a
cell index ${\bf i}=(i_a,i_b)$. The cell-Hamiltonian consists of a
kinetic energy term $t_a$ and the local interaction $U_d\cong 4$
eV\cite{Pen97}. 
In $H_d$ we droped a shift of the single particle levels 
which is only relevant for total energy considerations\cite{note1}.
The hopping between cells is defined by the last term, where
\begin{equation}
t_{\bf ij} = \left\{
\begin{array}{r l}
    t_b    & \mbox{for } \alpha=\beta,\; j_a=i_a,\; j_b=i_b\pm 1 \\
    t_{xy} & \mbox{for } \alpha\neq\beta,\; j_a=i_a\pm 1,\; j_b=i_b\pm \frac{1}{2}
\end{array}\right.
\end{equation}
Since the ratios between $U_d$ and the various hopping matrix elements
are quite large, we are confronted with a strong correlation problem.

\begin{figure}
\resizebox{0.40\textwidth}{!}{%
\includegraphics{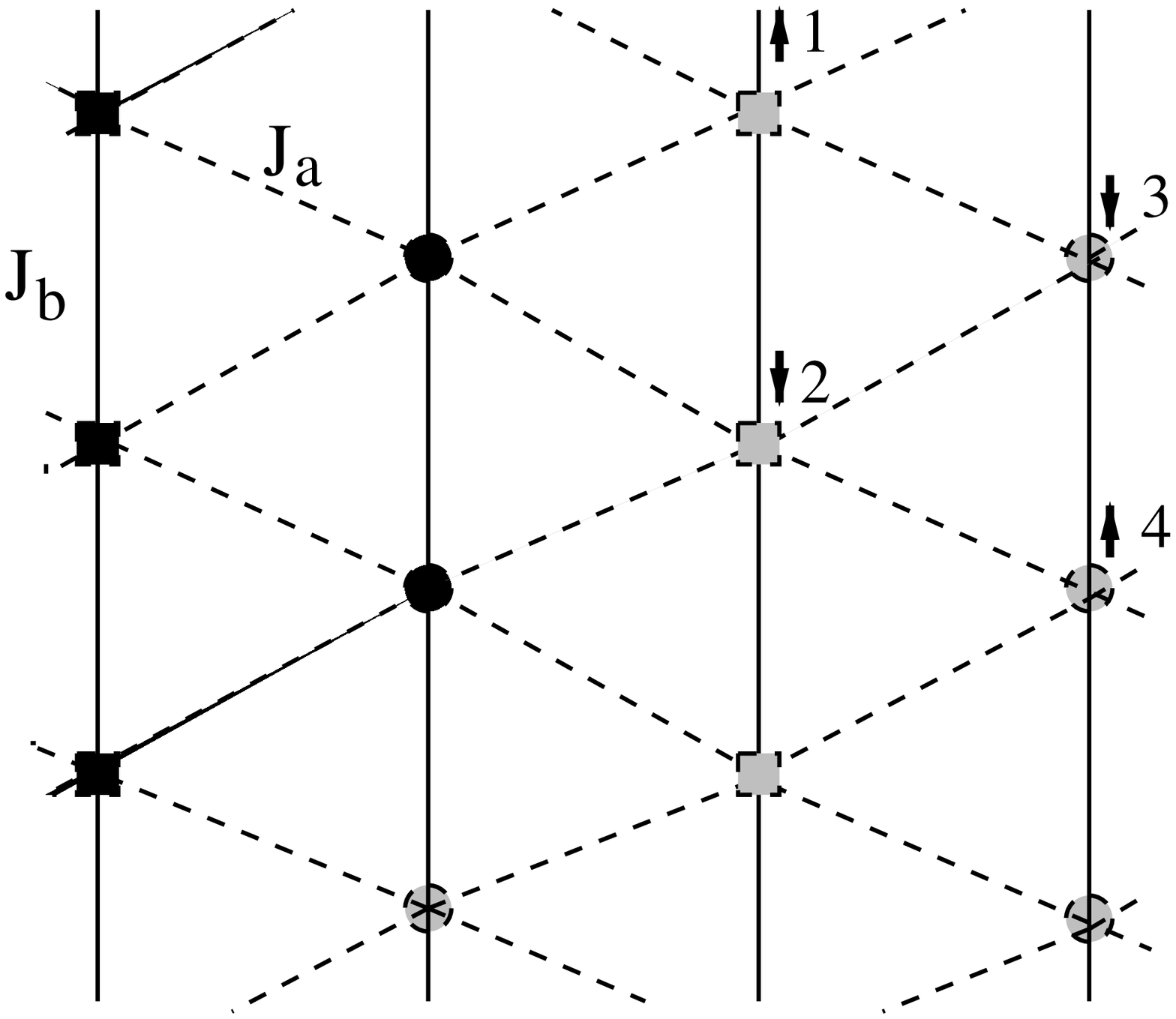}
}
\vspace{10mm}\noindent
\caption{Geometric structure of the effective spin model. 
A site corresponds to a V1-O3-V2 cell with a single $d$-electron.
Dark symbols indicate the part of the structure shown in Fig. 1.
The largest antiferromagnetic interaction is $J_b$, while the
frustrated interaction $J_a$ between neighboring b-chains is small.
The one-dimensional magnetic structure is further enhanced by the
topology of the lattice. The antiferromagnetic interaction $J_a$
of spin 3 with spins 1 and 2 on the neighbor chain is frustrated.
\label{fig2}}
\end{figure}

In the following we shall use a cell-perturbation method \cite{Jefferson92} 
based on the
above cell structure, where each V1-V2 cell contains in the
average a single $d$-electron. The advantage of the cell decomposition
is the capability to treat the local correlations exactly. The
complete set of states is then labeled by the quantum numbers of the
cells, i.e. including the number of electrons within a cell.
In the one electron sector the cell states are simply bonding and
antibonding states at energies $\pm t_a$ and corresponding operators
$b^{\dagger}_{{\bf i}\sigma}\;(a^{\dagger}_{{\bf i}\sigma})\;=\frac{1}{\sqrt{2}}
(d^{\dagger}_{{\bf i}1\sigma}\pm d^{\dagger}_{{\bf
    i}1\sigma})$, respectively. The low energy configurations 
$b^{\dagger}_{{\bf i}\sigma}b^{\dagger}_{{\bf j}\sigma'}$ of two
electrons in neighbor cells ${\bf i}$ and  ${\bf j}$ are coupled in
second order due to the hopping $t_{\bf ij}$. The intermediate states
have two electrons in one cell.  The 
low-energy singlet and triplet states have
excitation energies $\Delta E_s=2t_a -\frac{1}{2}(\sqrt{U_d^2+16t_a^2}-U_d)$
and $\Delta E_t=2t_a$, respectively.
Other singlet states are at much higher energy $E\geq U_d+2t_a$.
The Heitler-London singlet-triplet splitting of the low-energy
intermediate states turns out to be crucial for the
anisotropic nature of the magnetic properties.

The coupling of the $b^{\dagger}_{{\bf i}\sigma}b^{\dagger}_{{\bf j}\sigma'}$
configurations may be expressed in compact form
by the spin-1/2 Hamiltonian
\begin{equation}
H=\sum_{\langle {\bf i j} \rangle}\bigl[J_{\bf ij}^s\bigl({\bf S}_{\bf i}
 {\bf S}_{\bf j}-\frac{1}{4}n_{\bf i}n_{\bf j}\bigr) 
 -J_{\bf ij}^t\bigl({\bf S}_{\bf i} 
 {\bf S}_{\bf j}+\frac{3}{4}n_{\bf i}n_{\bf j}\bigr)\bigr],
\label{eq:Hspin}
\end{equation}
where ${\bf S_i}=\frac{1}{2}b^{\dagger}_{{\bf i},\sigma}
{\tau}_{\sigma \sigma'} b_{{\bf i},\sigma'}$ defines the spin of
bonding electrons in terms of the vector of Pauli spin matrices ${\bf \tau}$
and $n_{\bf i}=\sum_{\sigma}
b^{\dagger}_{{\bf i},\sigma} b_{{\bf i},\sigma}$ their density.
Here $J_{\bf ij}^s$ and  $J_{\bf ij}^t$ denote the coupling of
neighboring cells in the singlet- and triplet channel, respectively.
The relative size of the exchange integrals is strongly influenced by
the geometrical structure (Fig. 1 and 2).
The coupling of cells in a-direction, i.e. via $t_{xy}$,  yields 
$J_a^s=2(t_{xy}/2)^2/\Delta E_s$ and  $J_a^t=2(t_{xy}/2)^2/\Delta
E_t$. They differ only because of the singlet-triplet splitting
$E_{s-t}=\frac{1}{2}(\sqrt{U_d^2+16t_a^2}-U_d)$.
The coupling of cells in b-direction is
$J_b^s=2 t_{b}^2/\Delta E_s$ and  $J_b^t=0$, where the latter exchange
integral vanishes due to symmetry.
This has the effect that the total exchange constants in 
$H=\sum J_{\bf ij}{\bf S}_{\bf i}{\bf S}_{\bf j}$ almost cancel along
a-direction, i.e. $J_a=J_a^s-J_a^t$, whereas along the `b-chains'
$J_b=J_b^s$ there is no such reduction. 
These estimates are valid if the ratios $\frac{2 t_a}{U}$ and
$\frac{t'}{2 t_a}$, with $t'=\{t_b,t_{xy}/2\}$, are small compared to 1.
Numerical estimates for these exchange
integrals based on degenerate perturbation theory
are shown in Fig. 3 as function of the vanadium-oxygen (V1-O3)
hybridization $t_{pd}$.  
For the estimated value  $t_{pd}\simeq 1.2$ eV one finds $J_b\simeq 75$ meV
and $J_a\simeq 13$ meV.

\begin{figure}
\resizebox{0.40\textwidth}{!}{%
\includegraphics{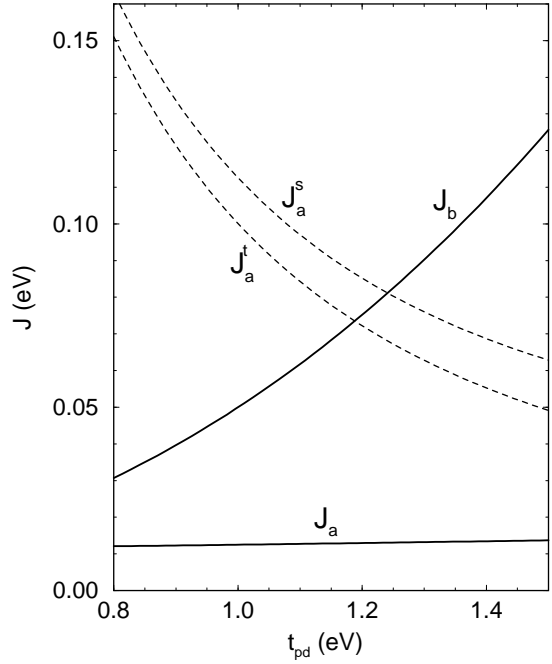}
}
\vspace{5mm}
\noindent
\caption{AF-exchange coupling between d-electrons in nearest 
  neighbor V1-O3-V2 bonds as
  function of $t_{pd}$: (a) $J_b$ along b-direction (solid line),
  (b) the small coupling $J_a=J_a^s-J_a^t$ between different b-chains mediated by
  $t_{xy}\sim 0.3$ eV results from the cancellation of the contributions 
  $J_a^s$ and $J_a^t$ from the
  Heitler-London split singlet and triplet intermediate states
  (dashed lines).
\label{fig:fig3}}
\end{figure}

Experimental estimates for the exchange constants are usually obtained
from the position of the maximum of the uniform susceptibility
$\chi(T)$.
In the following we study the effect of the interchain coupling 
$J_a$ and the thereby introduced frustration of the spin-1/2
model (3) using a finite temperature diagonalization
technique\cite{Jaklic94}. 
The geometrical structure of the lattice (Fig. 2) is similar to the 
resonating valence bond systems studied by Anderson and Fazekas
\cite{Anderson73}.
Results for a two-leg ladder (`railroad-trestle')  
with periodic boundary
conditions along a- and b-direction and different interchain coupling
strength $J_a/J_b$ are shown in Fig. 4 and 5. Since
the interchain coupling is frustrated the change of $\chi(T)$ is
relatively small.
The maximum of $\chi(T)$ is at the temperature
$T_{\chi}^{max}=a_{\chi}J_b$, where $a_{\chi}= 0.8$  for a $2\times 12$
system in the absence of the interchain coupling, i.e. $J_a/J_b=0$.
The exact result for the thermodynamic limit recently obtained
by Eggert {\it et al.}\cite{Eggert94}
using the thermal Bethe ansatz is $a_{\chi}= 0.6$. 
Interchain coupling leads to a small shift of the maximum by about 2\% to lower
temperatures for the value $J_a/J_b\sim
0.2$ estimated above (Fig. 4). 

\begin{figure}
\resizebox{0.40\textwidth}{!}{%
\includegraphics{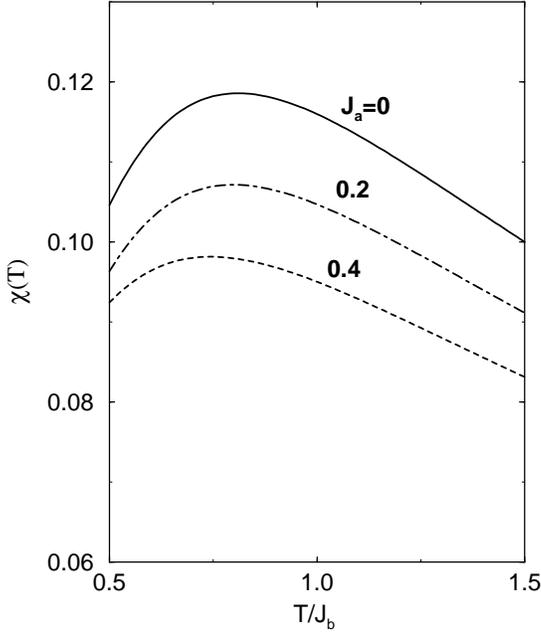}
}
\vspace{5mm}
\noindent
\caption{Uniform susceptibility $\chi(T)$ for a two-leg ladder
  ($2\times 12$) with the structure given in Fig. 2 for different
  interchain coupling strength $J_a/J_b = 0, 0.2$ and $0.4$ and
  periodic boundary conditions.
\label{fig:fig4}}
\end{figure}
\begin{figure}
\resizebox{0.40\textwidth}{!}{%
\includegraphics{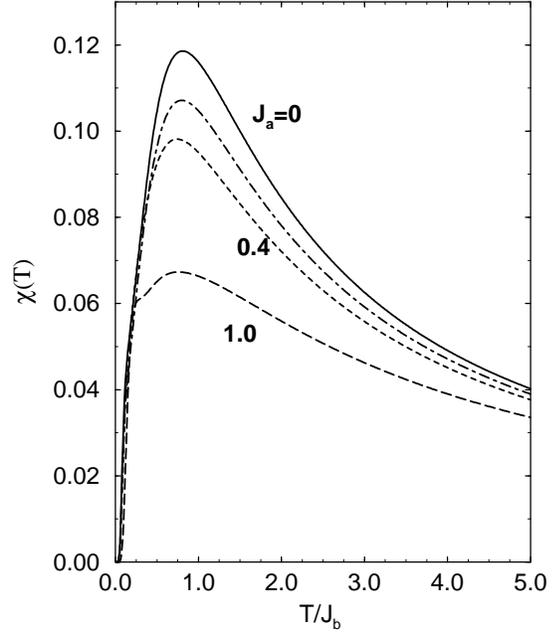}
}
\vspace{5mm}
\noindent
\caption{Uniform susceptibility $\chi(T)$ for a $2\times 12$  
  two-leg ladder for different
  interchain coupling strength $J_a/J_b = 0, 0.2, 0.4$ and $1.0$ on
  a large temperature scale. 
  The rapid drop below $T/J_b\leq 0.2$ is a finite size effect.
\label{fig:fig5}}
\end{figure}

We note that lattice fluctuations are expected to lead to a further
weakening of antiferromagnetism and an additional shift of the maximum of
$\chi(T)$ to lower $T$.  A recent study of this effect by Sandvik {\it
  et al.}\cite{Sandvik97} shows that this may lead to a 
reduction of $T_{\chi}^{max}$ by 15\% for a one-dimensional Heisenberg
chain. Combining these two effects one arrives at the estimate
$a_{\chi}^{tot}\sim 0.50$. From the susceptibility 
measurements\cite{Isobe96,Mila96} 
for  $\alpha'$-NaV$_2$O$_5$  one finds 
$T_{\chi}^{max}\sim 350$ K, which leads to the estimate
$J_b^{exp}\sim 700$ K. This is in reasonable agreement with the theoretical
value $J_b\sim 75$ meV derived in this work.
In view of the simplicity of the model and the
rough estimates of the parameters this is quite satisfactory.

We briefly comment on the band picture, i.e. ignoring the effect of
electron correlations. Since double-exchange is not contained in this
scheme, the only large hopping matrix element is $t_{xy}$. This
leads to two degenerate one-dimensional $d_{xy}$-bands from the
zig-zag chains V2-V1$'$-V2-V1$'$ and V2$'$-V1-V2$'$-V1 (Fig. 1), respectively,
since there are two zig-zag b-chains per unit cell. 
Each band is quarter-filled and one may expect a usual Peierls
transition to occur due to an appropriate dimerization of the
structure in b-direction. Although this could explain the insulating properties
at room temperature, such an explanation is obviously incorrect,
since it would rule out a spin-Peierls transition at low temperature. 
Of course the large ratio $U_d/t_{xy}$ already excludes the straightforward
application of the band picture.
Nevertheless for sufficiently large $t_{xy}$ one expects that the correlated 
band picture applies. Estimates of the ground state energies 
for the localized regime and the band picture suggest that this is the case 
for $t_{xy}>1.5 t_a$.

The bands obtained by a real bandstructure calculation deviate from this
idealistic picture due to small interchain couplings, which is also
reflected in shorter V1-O3-V2 bonds\cite{Grin97}. The full bandstructure
calculation predicts a metallic state.

We remark that photoemission experiments in combination with XAS would
provide a sensitive test of the present picture, since the lowest
unoccupied states are the singlet and triplet states, while higher
lying `two-particle' states 
are split by $U_d$. Optical conductivity measurements
on the other hand should show a relatively small `single particle' gap
of order $2 t_a$.

In summary we have shown that $\alpha'$-NaV$_2$O$_5$ is an insulator
due to strong correlations, i.e. large $U_d$. However since the single
electron orbitals have bonding character, i.e. involving two V-atoms,
the magnetic structure is induced via 2-electron molecular singlet and
triplet states which are Heitler-London split. 
Due to the almost perfect cancellation of the 
triplet- and singlet-interactions in a-direction the spin system is
essentially one-dimensional. We stress that doubly occupied states
with energy $\sim U_d$, which usually contribute to 
the magnetic interaction in Mott-Hubbard
insulators, have little influence since they are at much higher energy.
Therefore we suggest the name {\it Heitler-London insulator} as a more 
precise characterization for such systems.

We acknowledge stimulating discussions with J. van den Brink, Yu. Grin, 
A. M. Oles and H.-G. von Schnering.



\begin{thebibliography}{}
\bibitem{note1}
Covalency effects lead to diagonal shifts of $-4 t_a$, $-2 t_a-t_a^U$ 
and $-2 t_a^U$ in the $n_d=0,1$ and 2 sectors, respectively, where 
$t_a^U\sim t_{pd}^2/(e_d-e_p+U_d)$. Although
these shifts are important for ground state energies, they do not affect
the excitation energies considered here.
\end{thebibliography}
\end{document}